\documentclass[aps, pra, twocolumn, showpacs,footinbib,superscriptaddress]{revtex4-1}
\usepackage{graphicx, subfigure,braket}
\usepackage{amsmath,amssymb}
\usepackage{color}
\usepackage{natbib,hyperref}
\usepackage{placeins}
\usepackage{qcircuit}

\begin{document}

\title{Nonadiabatic quantum control of quantum dot arrays with fixed exchange using Cartan decomposition}

\author{David W. Kanaar}
\affiliation{Department of Physics, University of Maryland Baltimore County, Baltimore, MD 21250, USA}
\author{Utkan G\"ung\"ord\"u}
\affiliation{Laboratory for Physical Sciences, College Park, Maryland 20740, USA}
\affiliation{Department of Physics, University of Maryland, College Park, Maryland 20742, USA}
\author{J.~P.~Kestner}
\affiliation{Department of Physics, University of Maryland Baltimore County, Baltimore, MD 21250, USA}

\begin{abstract}
In semiconductor spin qubits which typically interact through short-range exchange coupling, shuttling of spin is a practical way to generate quantum operations between distant qubits. Although the exchange is often tunable through voltages applied to gate electrodes, its minimal value can be significantly large, which hinders the applicability of existing shuttling protocols to such devices, requiring a different approach. In this work, we extend our previous results for double- and triple-dot systems, and describe a method for implementing spin state transfer in long chains of singly-occupied quantum dots in a nonadiabatic manner.
We make use of Cartan decomposition to break down the interacting problem into simpler problems in a systematic way, and use dynamical invariants to design smooth nonadiabatic pulses that can be implemented in devices with modest control bandwidth.
Finally, we discuss the extensibility of our results to directed shuttling of spin states on two-dimensional lattices of quantum dots with fixed coupling.
\end{abstract}

\maketitle

\section{Introduction}

Spins confined in semiconductor quantum dots are a promising platform for the realization of a useful quantum computer. One-qubit gate fidelities around 99.9\% fidelities are routinely achieved in Si/SiO$_2$ \cite{Veldhorst2014,Yang2018a} and Si/SiGe heterostructures \cite{Yoneda2017}, and two-qubit gate fidelities nearing 99\% for entangling neighboring qubits via Heisenberg exchange interaction have recently been reported in \cite{Xue2022}.
Entangling qubits that are separated over long-distances can be achieved via spin-photon coupling \cite{Mi2017,Warren2019,Benito2019,Warren2021,Harvey-Collard2021} by coupling the dots to  to a common microwave cavity and selectively bringing the qubits in and out of resonance. For intermediate length scales, coherent shuttling of spins via adiabatic passage \cite{Ginzel2020,Mills2019,Nakajima2018,Noiri2022,Fujita2017,Baart2016,Seidler2021,Flentje2017,Edlbauer2021,Jadot2021,Mortemousque2021,Zwerver:2022oac} is a viable route to entangle distant qubits, which can be achieved by modulating the voltages applied in an array of empty quantum dots to the gate electrodes without the overhead of a resonator. An alternative method of transferring qubit information is through repeated SWAP operations while maintaining the states of all other qubits. SWAP gates have been experimentally been implemented on two qubit devices \cite{Sigillito2019a,Petit2020,vanRigglelen2022}. Ref.~\cite{Gullans2020} describes conditional spin state transfer in a linear array of filled quantum dots using pulsed control of the exchange coupling. Our work describes a method of using repeated SWAP gates in a 1D or 2D array of qubits whose exchange interaction is constant in time. 

Heisenberg exchange interaction can allow direct implementation of SWAP gates, provided that it can be turned on and off between the neighboring quantum dots. However, when there is a large variation in energy splitting between adjacent qubits, the Heisenberg coupling effectively reduces to an Ising $ZZ$ coupling in the rotating wave approximation, thus preventing the direct swapping of spins without application of local driving fields. Furthermore, in devices with an always-on exchange coupling, implementation of any quantum gate can be a challenge due to crosstalk \cite{Huang2019b}. Nonetheless, in certain scenarios, such systems with high connectivity can be systematically broken into a simpler set of noninteracting subsystems (which may however share control degrees of freedom) by means of Cartan decomposition \cite{Gungordu2012a}, as exemplified in \cite{Gungordu2019b,Gungordu2020b,Gungordu2020a} in quantum double dots and in \cite{Kanaar2021} for triple dots. However, these methods are not readily applicable to networks of quantum dots with larger number of spins, or topologies beyond a one-dimensional chain.

In this paper, we extend the earlier works to arbitrary long chains ($>3$) of quantum dots with always-on couplings, and present a method of achieving spin state transfer in a nonadiabatic manner using Cartan decomposition, by using iSWAP as the primitive. We also discuss how these results might be extended to two-dimensional lattices. This paper is organized as follows. We first give a brief review of Cartan decomposition in the context of dynamical invariants. In Sec.~\ref{sec:Model4} that follows, a model describing a chain of singly loaded quantum dots that is driven by an ESR line is given, operating in a regime that is specified. Sec.~\ref{sec:iswapPLUSindividualgates} starts by showing how to build the iSWAP out of these components. Then it continues by showing how to create the elementary gates of the iSWAP gate using square pulse sequences. Finally, in Sec.~\ref{sec:dynamicalinvariants} the individual gates are partially implemented using smooth pulses with modest bandwidth requirements, by parameterizing the problem using dynamical invariants.

\section{Cartan decomposition and dynamical invariants}
\label{sec:cartan}
A dynamical invariant $I(t)$ is an operator whose expectation value is conserved, and obeys the defining equation
\begin{align}
 i \partial_t I(t)  = [H(t), I(t)]
 \label{eq:DI}
\end{align}
where $H(t) \in  \mathfrak{su}(N)$ is the Hamiltonian of the system. This is equivalent to choosing to work in the Heisenberg picture with an operator, $I^{(H)}(t)$, whose total time derivative is $0$ \cite{Gungordu2012a}. It allows expressing the time-evolution operator as
\begin{align}
 U(t;0) = \sum_n e^{i \alpha_n(t)} \sum_i |\phi_n(t)\rangle \langle \phi_n(0) |,\\
 \qquad \alpha_n(t) = \int_0^t ds \langle \phi_n(s)| \left(  i \partial_s - H(s) \right) |\phi_n(s) \rangle
\end{align}
where $|\phi_n(t)\rangle$ are  the instantaneous eigenvectors of $I(t)$ and $\alpha_n(t)$ are the associated Lewis-Riesenfeld phases \cite{LewisRiesenfeldPhase}. The eigenvalues of $I(t)$ are constants in time. One useful aspect of dynamical invariants is that they provide a set of states that are transitionless, such that if the system is initialized to  $|\phi_m(0)\rangle$, it will later evolve into $e^{i \alpha_m(t)} |\phi_m(t) \rangle$. This form is similar to that of adiabatic evolution of the eigenvectors of $H(t)$, except it is valid in the nonadiabatic regime as well \cite{ChenMuga2011}. The two terms in the Lewis-Riesenfeld phase can be recognized as the geometric and dynamical phase, which makes the dynamical invariant a natural parametrization to study geometric quantum gates \cite{Gungordu2014,Colmenar2021,Ralph2022dynamical}.

In quantum computing, Lie algebras are useful tools to understand the controllability of a system \cite{DAlessandro2021}. The Lie algebra generates its corresponding Lie group via the exponential map, and is a vector space. In unitary quantum dynamics, Hamiltonians are described by Lie algebras, and the unitary time-evolution operator belongs to the corresponding Lie group. Cartan decomposition of Lie algebras is a useful tool to relate the set of possible quantum gates and control schemes for a given system, or systematically decompose the quantum evolution into local and non-local parts \cite{DAlessandro2021}. The Cartan decomposition of a Lie algebra $\mathfrak g = \mathfrak k \oplus \mathfrak p$ satisfies the following relations
\begin{align}
[\mathfrak{k}, \mathfrak{k}] \subseteq \mathfrak{k}, \quad [\mathfrak{k}, \mathfrak{p}] \subseteq \mathfrak{p}, \quad [\mathfrak{p}, \mathfrak{p}] \subseteq \mathfrak{k}\label{eq:Cartandecomp1}
\end{align}
In Eq.~\eqref{eq:Cartandecomp1} $\mathfrak k$ is a Lie subalgebra of $\mathfrak g$, and any subalgebra of $\mathfrak p$ is Abelian. These relations have a direct implication on Eq.~\eqref{eq:DI}: the dynamical invariant can be chosen to be in the same subalgebra $\mathfrak{k}$ as $H(t)$, or in the vector space $\mathfrak{p}$ \cite{Gungordu2012a}, where the possible choices for $\mathfrak{k}$ are determined by the maximal subalgebras of the Lie algebra $\mathfrak{su}(N)$. 

The minimal interesting example is the case of two-qubits for which $N=4$, whose maximal subalgebras are tabulated in Table \ref{tab:embeddings}.
As we will show in the next section, a pair of spin qubits in a double quantum dot can be described using the following rotating frame Hamiltonian,
\begin{align}
    H = \frac{J}{4} Z_1 Z_2 + \sum_{i=1}^2 \frac{\Omega_i}{2} (\cos (\phi_i)X_{i}+\sin (\phi_i)Y_{i}),
\end{align}
 where $\sigma_{Z,i}$ is the $Z$ Pauli operator on the $i$-th qubit, $J$ is the strength of the exchange coupling, and $\Omega_i, \phi_i$ are the amplitude and phase of the drive. When the qubits are driven one at a time, say $\Omega_2=0$ for concreteness, the Hamiltonian fits into the $\mathfrak{su}(2) \oplus \mathfrak{su}(2) \oplus \mathfrak{u}(1)$ with the Cartan decomposition

\begin{align*}
\mathfrak{su}(4) =\text{span}(\left\{ X_1 \frac{1+ Z_2}{2}, Y_1 \frac{1+ Z_2}{2}, Z_1 \frac{1+ Z_2}{2} \right\} \sqcup\\
\left\{ X_1 \frac{1- Z_2}{2}, Y_1 \frac{1- Z_2}{2}, Z_1 \frac{1- Z_2}{2} \right\} \sqcup \\
\underbrace{\{Z_2 \}}_{Q} \sqcup \underbrace{\{ X_1 Y_2, Y_1 Y_2, Z_1 Y_2, X_2 \}}_{P} \sqcup \\
\underbrace{\{ X_1 X_2, Y_1 X_2, Z_1 X_2, -Y_2 \}}_{\bar P})
\end{align*}

We remark that $P$ and $\bar P$ are related to each other by an infinitesimal $\mathfrak{u}(1)$ ``charge conjugation" \cite{Gungordu2012a}
\begin{align}
    [Q, P] = -2i \bar P, \qquad [Q, \bar P] = 2 i P.
\end{align}
Since $\mathfrak{su}(2) \oplus \mathfrak{su}(2)$ portion of $\mathfrak{su}(2) \oplus \mathfrak{su}(2) \oplus \mathfrak{u}(1)$ does not mix $P$ and $\bar P$ under commutation, when the Hamiltonian is missing the $\mathfrak{u}(1)$ term,
it becomes possible to choose a dynamical invariant in span of $P$ or $\bar P$. As a result, in the 15-dimensional real adjoint representation, the Hamiltonian becomes block diagonal with the structure $3+3+1+4+4$ \cite{Gungordu2012a}.

A trivial example is the case of uncoupled driven qubits with $J=0$ and $\Omega_i \neq 0$. In this case, the Hamiltonian belongs to the maximal Lie subalgebra $\mathfrak{su}(2) \oplus \mathfrak{su}(2)$ given by the span of $\{X_1, Y_1, Z_1\} \sqcup \{X_2, Y_2, Z_2\}$, and its adjoint representation has the block diagonal structure $3 + 3 + 9$.

The above arguments for dynamical invariants can also be made for the density matrix $\rho(t)$ which obeys the same equation as dynamical invariants, although differs from $I(t)$ by an unimportant (in this context) term that is proportional to identity matrix as required by the trace condition $\text{Tr}(\rho) = 1$.

\begin{table}
\begin{center}
\begin{tabular}{| l | c |}
\hline
$\mathfrak{su}(3) \oplus \mathfrak u(1)$ & $8 + 1 + 3 + 3$ \\ 
$\mathfrak{su}(2) \oplus \mathfrak{su}(2)$ & $3 + 3 + 9$ \\
$\mathfrak{su}(2) \oplus \mathfrak{su}(2) \oplus \mathfrak u(1)$ & 3 + 3 + 1 + 4 + 4 \\ 
$\mathfrak{so}(5)$ & $10 + 5$ \\
\hline
\end{tabular}
 \caption{Maximal subalgebras of $\mathfrak{su}(4)$  \cite{Slansky1981} and the corresponding partitioning of the 15-dimensional vector space in the adjoint representation, which can be used to determine the block diagonal structure of the Hamiltonian in the adjoint representation as listed in the right column.  $\mathfrak{p}$ is partitioned into two for maximal subalgebras containing a $\mathfrak{u}(1)$.}
 \end{center}
\label{tab:embeddings}
\end{table}

\section{Model}\label{sec:Model4}
\begin{figure}
    \centering
    \includegraphics[width=0.8\linewidth]{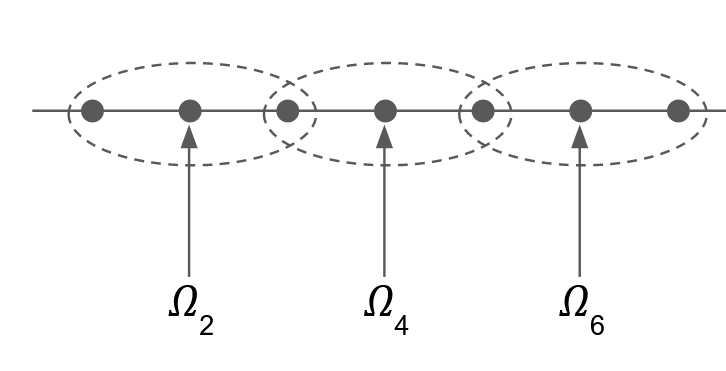}
    \caption{Schematic representation of the driving on a chain of qubits which results in $(N-1)$ $\mathfrak{su}(2)$s. The dashed circle indicate the qubits involved in the separate $\mathfrak{su}(2)$s.}
    \label{fig:0schematic}
\end{figure}

A chain of $N$ nearest-neighbor exchange coupled silicon quantum dots in the presence of magnetic fields is well known to be described by the Heisenberg model Hamiltonian \cite{Loss1998}
\begin{equation}
    H=\sum_{i=1}^{N-1} J_{i,i+1} \left(\mathbf{S}_{i}\cdot \mathbf{S}_{i+1}-\frac{1}{4}\right) +\sum_{i=1}^{N} \mu g_{i} \mathbf{B}_{i} \cdot \mathbf{S}_{i}
    \label{eq:H14}
\end{equation}
where $J_{i,i+1}$ is the exchange strength between neighboring dots, $\mathbf{S}_i$ is the spin operator of the $i$-th qubit, $\mu$ is the Bohr magneton, $g_i$ is the $g$-factor of the $i$-th qubit and $\mathbf{B}_i$ is the magnetic field at the $i$-th qubit. The exchange coupling in this paper is taken to be fixed, always-on, and the same between each qubit meaning $J_{i,i+1}=J$. Additionally, in this work the magnetic field is applied only in the $Z$ and $X$-directions and will now be described by the Zeeman energy  $E_{z,i}=\mu g_i B_{z,i}$ for the $Z$-direction and the envelope, $\Omega_i$, frequency, $\omega_i$, and phase $\phi_i$, $\Omega_i \cos(\omega_i t +\phi_i)=\mu g_i B_{x,i}$ for the $X$-direction. The Zeeman energies are taken to be large and constant with the differences between Zeeman energies of neighboring qubits being large compared to the other terms in the Hamiltonian as has been the case in some experiments \cite{Yoneda2017,Hendrickx2021,Veldhorst2015}. To avoid the large Zeeman energies, the Hamiltonian in \eqref{eq:H14} is moved to the rotating frame $R=e^{\frac{i}{2} \sum_i^N E_{z,i} t \sigma_{Z}^{(i)}}$ As a result of the large difference in Zeeman energies it is possible to apply the rotating wave approximation to the Hamiltonian in the rotating frame $H_R=R H R^{\dagger}+i \hbar (\partial_t R) R^{\dagger}$ which results in:
\begin{equation}
    H_{R,N}\approx\sum_{i=1}^{N-1} \frac{J}{4}Z_{i}Z_{i+1}+\sum_{i=1}^{N} \frac{\Omega_i}{2} (\cos (\phi_i)X_{i}+\sin (\phi_i)Y_{i})
    \label{eq:HRN}
\end{equation}
If in Eq.~\eqref{eq:HRN} every other qubit is locally driven, i.e., $\Omega_i=0$ for even or odd $i$ as represented schematically in Fig.\ref{fig:0schematic}, then there exists a decomposition of the Hamiltonian in terms of $N-1$ $\mathfrak{su}(2)$s and $N-1$ $\mathfrak{u}(1)$s within this $\mathfrak{su}(2^N)$ algebra. This can be shown by examining a three qubit string within this Hamiltonian such as $H_{R,3}$ with $\Omega_i=\Omega(t) \delta_{i,2}$.
\begin{equation}
    H_{R,3}= \frac{J}{4} (Z_{1}Z_{2}+Z_{2}Z_{3})+\frac{\Omega(t)}{2} (\cos (\phi(t))X_{2}+\sin (\phi(t))Y_{2}) )
    \label{eq:HR3}
\end{equation}
As was shown in Ref.~\cite{Kanaar2021,Gullans_2019} this Hamiltonian can be decomposed into four possible $\mathfrak{su}(2)$ $H_{\pm,\pm}$
\begin{equation}
    H_{R,3} = H_{++}+H_{+-}+H_{-+}+H_{--}
\end{equation}
where, taking the phases of the driving $\phi_i = 0$,
\begin{align}
     H_{++} &= \frac{J}{2}Z_{++}+\frac{\Omega}{2} X_{++} \label{eq:Hpp2}\\
     H_{+-} &= -\frac{\Omega}{2} X_{+-} \label{eq:Hpm2}\\
     H_{-+} &=  -\frac{\Omega}{2} X_{-+} \label{eq:Hmp2}\\
     H_{--} &= -\frac{J}{2}Z_{--}+\frac{\Omega}{2} X_{--} \label{eq:Hmm2}
\end{align}
and
\begin{align}\label{eq:eff_Z}
    Z_{s_1 s_2} &=\frac{1}{4} \left(Z_{1}+s_1 I\right) Z_{2}  \left(Z_{3}+s_2 I\right) \\ \label{eq:eff_X}
    X_{s_1 s_2} &= \frac{1}{4}\left(Z_{1}+s_1 I\right) X_{2}  \left(Z_{3}+s_2 I\right),
\end{align}
with $s_i \in \{+,-\}$. $H_{+-}$ and $H_{-+}$ in Eqs.~\eqref{eq:Hpm2} and \eqref{eq:Hmp2} have been further simplified to $\mathfrak{u}(1)$s as a result of choosing $J_{i,i+1}=J$. \par
\begin{figure}[h!]
    \centering
\Qcircuit @C=1.0em @R=1.0em { 
&\gate{Z_{-\frac{\pi}{2} }} & \gate{ \mathbf{H} } & \multigate{1}{ ZZ_{ \frac{\pi}{2} } } &  \gate{Z_{\frac{\pi}{2} }} & \gate{ \mathbf{H} } & \multigate{1}{ ZZ_{ \frac{\pi}{2} } } &  \gate{Z_{\frac{\pi}{2} }} & \gate{ \mathbf{H} } & \qw\\
& \gate{Z_{-\frac{\pi}{2} }} & \gate{ \mathbf{H} } & \ghost{ ZZ_{ \frac{\pi}{2} } } &  \gate{Z_{\frac{\pi}{2} }} & \gate{ \mathbf{H} } & \ghost{ ZZ_{ \frac{\pi}{2} } } &  \gate{Z_{\frac{\pi}{2} }} & \gate{ \mathbf{H} } & \qw  }
    \caption{ Circuit diagram of the gates which result in an iSWAP gate between two qubits. $\mathbf{H}$ denotes a Hadamard gate and $G_{\theta} = \exp{\left(-i\frac{\theta}{2} G\right)}$.}
    \label{fig:1qubits}
\end{figure}
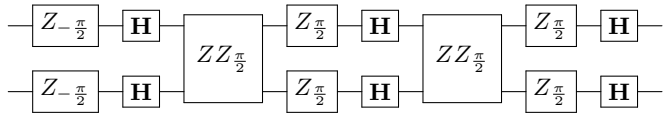
\section{Sequence for doing an iSWAP gate using  square pulses}\label{sec:iswapPLUSindividualgates}
\subsection{Composition of an iSWAP gate} \label{sec:iswap}
An iSWAP gate can be created from Hadamard gates, $Z$ rotations and $\frac{\pi}{2}$ $ZZ$ rotations using the scheme designed with methods from Ref.~\cite{Qiskit,NielsenandChuang} and shown in Figure~\ref{fig:1qubits}. The $Z$ rotations will be done virtually \cite{Mckay2017}, the Hadamard will be performed by virtual $Z$-gates and a $\frac{\pi}{2}$ $X$ rotation and the $ZZ$ rotations will be performed using a sequence of $X$ gates and turned off driving.

In Fig.~\ref{fig:1qubits} the individual Hadamard gates are not performed simultaneously because this approach relies on the decomposition of the Hamiltonian into $\mathfrak{su(2)}$s which does not work when driving neighboring qubits simultaneously.
There are a total of six Hadamard gates and two $\frac{\pi}{2}$ $ZZ$ rotations, which we will show below takes a total time of $t_\text{iSWAP}\approx 116 /J$ using square pulses. 

\subsection{Individual gates}\label{sec:Individualgates}
Within an SU(2) it is possible to do any rotation using the Euler angle decomposition which only requires rotations around two perpendicular axes. This is possible for the SU(2)s generated from Eq.~\eqref{eq:Hpp2} and \eqref{eq:Hmm2} by applying $\Omega=\pm J$ in alternating fashion. For the $\mathfrak{su}(2)$s that simplify into $\mathfrak{u}(1)$s in Eq.~\eqref{eq:Hpm2}-\eqref{eq:Hmp2} it is only possible to do one type of rotation, however we will show that this is sufficient for doing an iSWAP gate.

\subsubsection{$\frac{\pi}{2}$ $X$ rotation concurrently with identities in other subspaces}\label{subsec:Xgate}
The Pauli $X$ operator on the second qubit is equivalent to an equally weighted sum of $X_{s_1 s_2}$ generators, 
\begin{equation}
    X_{2} = \left(X_{++}-X_{+-}-X_{-+}+X_{--}\right).
    \label{eq:XinXs1s2}
\end{equation}
This means a $\frac{\pi}{2}$ rotation around $X$ on the second qubit is equivalent to performing an $s_1 s_2 \frac{\pi}{2}$ rotation around each of the $X_{s_1 s_2}$ generators.
The $++$ and $--$ Hamiltonians are symmetric, therefore the $X_{++}$ and $X_{--}$ rotations can be achieved at the same time. The desired $\frac{\pi}{2}$ $X_{++}$ and $X_{--}$ rotations are created with a pulse sequence which implements the Euler decomposition by switching  $\Omega$ between $\pm J$ and using
 \begin{align*}
     e^{-i \frac{J t_1}{2}\left(Z_{++}+X_{++}\right)}  e^{-i \frac{J t_2}{2}\left(Z_{++}-X_{++}\right)} \times\\
    e^{-i \frac{J t_3}{2}\left(Z_{++}+X_{++}\right)} = e^{-i\frac{\pi}{4}X_{++}}.
 \end{align*}
The solution with the minimum elapsed time is
 \begin{align}
     t_1=t_3&=\frac{\arctan(1/\sqrt{2})}{\sqrt{2}J}, \\ t_2&=\frac{5\sqrt{2}\pi }{3J}
 \end{align}
At the same time evolution in the $+-$ and $-+$ subspaces also happens, but this is not generally equivalent to the intended $-\frac{\pi}{2}$ rotation. To create the desired  $-\frac{\pi}{2}$ rotation in these subspaces a final step is added to the sequence. This step ensures that the pulse area under $\Omega \left(t\right)$ for the entire sequence is equivalent to a $-\frac{\pi}{2}$ in the $+-$ and $-+$ spaced. Additionally, the rotation in the $++$ and $--$ subspaces is preserved by producing an identity up to $\pi$ $Z_1$ and $Z_3$ rotations, which can be compensated virtually through the phase degree of freedom\cite{Knill2000,Mckay2017}. These conditions can be expressed as
 \begin{align}
     \frac{J}{2}(t_1-t_2+t_3) + \frac{\Omega}{2}t_4 &=2 m \pi+ \frac{\pi}{4}, \\
     t_4 \sqrt{\left( \frac{J}{2}\right)^2+\left(\frac{\Omega}{2}\right)^2} &=n \pi.
 \end{align}
for any integers $n$ and $m$.
Then the solution for the length and drive amplitude of the final time step that gives a real solution in this case is
\begin{align}
     \Omega &= \frac{J(\pi-2 J \tau)}{\sqrt{\left(9\pi-2J\tau\right)\left(7\pi+2J \tau\right)}} \approx 0.991 J,
     \\
     t_4 &= \frac{\sqrt{\left(9\pi-2J\tau\right)\left(7\pi+2J \tau\right)}}{2J} \approx 8.926/J,
\end{align}
where $\tau = t_1-t_2+t_3$. The total time of the $X$ rotation is $t_{\frac{\pi}{2}X} \approx 17.94/J$. For a chain of $N$ qubits it is also necessary to create identities in this structure of $\mathfrak{su}(2)$s on all other qubits at the same time. This can be achieved by a two part pulse in which $\Omega_1=-\Omega_2$ such that the $+-$ and $-+$ spaces create an identity. The identity in the $++$ and $--$ spaces is found by driving at strength $\Omega=2\sqrt{(2 n \pi/t)^2-(\frac{J}{2})^2}$ for any pulse time $t$ and integer $n$ that gives a real solution. Additionally, any necessary virtual $Z$ rotations can be done during any of the $X$ rotations sequences mentioned in this section by adjusting the phase, $\phi_i$,
\begin{equation}\label{eq:virtualZ}
    e^{- i \theta (\cos (\phi_i)X_i+\sin (\phi_i)Y_i)}=e^{- i \frac{\phi_i}{2 }Z_i}e^{- i \theta X_i}e^{ i \frac{\phi_i}{2 }Z_i}.
\end{equation} 
\subsubsection{Hadamard gates and virtual $Z$ rotation}
It is possible to create a Hadamard gate, $\mathbf{H}$, up to a global phase by doing a combination of a $\frac{\pi}{2}$ $X$ rotation and two $Z$ rotations,
\begin{equation}
    \mathbf{H_i}=e^{- i \frac{\pi}{4 }Z_i}e^{- i \frac{\pi}{4 }X_i}e^{- i \frac{\pi}{4 }Z_i}.
    \label{eq:HadamardGate}
\end{equation}

At this point we note that for spin state t in this scheme only $X$ rotations, $ZZ$ rotations and $Z$ rotations are necessary and because $ZZ$ rotations commute with the $Z$ rotations it will always be possible to resolve the necessary $Z$ rotations using virtual-$Z$ rotations generated during the $X$ rotations.

\subsubsection{$\frac{\pi}{2}$ ZZ rotation  concurrently with identities in other subspaces}\label{subsubsec:ZZrot}
The $Z_{1}Z_{2} $ operator is equivalent to a weighted sum of $Z_{s_1 s_2}$,
\begin{equation}
    Z_{1}Z_{2} = \left(Z_{++}+Z_{+-}-Z_{-+}-Z_{--}\right).
    \label{eq:ZinZs1s2}
\end{equation}
From the Hamiltonians in Eq.~\eqref{eq:Hpm2} and \eqref{eq:Hmp2} it is apparent that this cannot be directly created since the $Z_{-+}$ and $Z_{-+}$ generators are absent. It is, however, possible to acquire a phase on the $Z_{1}Z_{2} $ + $Z_{2}Z_{3}$ term in the exponential by simply turning off the $\Omega$ driving for a time $\pi/2J$, then performing an $X_{3}$ gate, then repeating those two steps to echo out the $Z_{2}Z_{3}$ term and produce the desired gate, $\exp{(-i \pi/4 Z_{1}Z_{2})}$. This echo works because $X_{3}$ commutes with $Z_{1}Z_{2}$ and anticommutes with $Z_{2}Z_{3}$.
For a larger chain it is necessary to do identities in all other subspaces at the same time, which is possible by using the same trick mentioned at the end of Sec.~\ref{subsec:Xgate} with the time being $t_{ZZ}=\pi/J$.

The $X$ gates required to produce the $\frac{\pi}{2}$ $ZZ$ rotation can be composed of two $\frac{\pi}{2}$ $X$ rotations as mentioned in Sec.~\ref{subsec:Xgate}. 
However, this rotation can be performed faster by using the four-step process of Sec.~\ref{subsec:Xgate} for a $\pi$ $X$ rotation directly. Using $\tau=t_1-t_2+t_3$ the shortest real solutions for an $X$ gate if found with the times and qubit driving strengths
 \begin{align}
     t_1&=t_3=\frac{\pi}{2 \sqrt{2} J},\\
     t_2&=\frac{7\pi}{2 \sqrt{2} J}, \\ t_4&=\frac{\sqrt{\left(5\pi-J\tau\right)\left(3\pi+J \tau\right)}}{J} \approx 9.07233/ J,\\
     \Omega_1&=\Omega_3=-\Omega_2=J,\\ \Omega_4&=\frac{J(\pi- J \tau)}{\sqrt{\left(5\pi+J\tau\right)\left(3\pi+J \tau\right)}} \approx 0.95843 J.
 \end{align}
This results in the total evolution operator, $U=\Pi_{j=1}^4 e^{i t_j\left( \frac{J}{4}\left(Z_{1}Z_{2}+Z_{2}Z_{3}\right)+\Omega_j X_2 \right)}$, being equivalent to an $X$ gate up to a global phase. The time for this $X$ gate is $t_{\pi X}\approx19.07/J$, which is almost twice as fast as two $\frac{\pi}{2}$ $X$ rotations. This makes the combined time for the $\frac{\pi}{2}$ $ZZ$ rotation to $\approx 41.28/J$.

However, in the context of the overall circuit of Fig.~\ref{fig:1qubits}, 
instead of echoing away $Z_2 Z_3$ within each of the entangling segments of the circuit (which requires four $X_3$ gates in total), we can save time by just performing an $X_3$ after each $\pi/2$ $ZZ$ rotation of Fig.~\ref{fig:1qubits}. Since qubits that are not nearest neighbors can safely be driven simultaneously without ruining the $\mathfrak{su(2)}$ decomposition of the Hamiltonian, the $X_3$ gate can be done at the same time as the trailing Hadamards on qubit 1, as long as the difference in times is accounted for with an additional identity on qubit 1 to pad the time as discussed in Sec.~\ref{subsec:Xgate}. Thus, the entire circuit requires six $\frac{\pi}{2}$ $X$ rotations and two $\pi$ $X$ rotations in addition to the instantaneous virtual $Z$ rotations and the two $\pi/2$ $ZZ$ rotations, and accounting for the fact that gates on nearest neighbors cannot be performed in parallel, the total time for the iSWAP using the above square pulses is $t_\text{iSWAP} = 4 t_{\frac{\pi}{2}X} + 2 t_{\pi X} + 2\pi/J \approx 116/J$.

\subsubsection{Rotations on edge qubits}
As an aside, we note that in the case where the driving is on a qubit at an end of the linear array, the decomposition mentioned above is not necessary because the Hamiltonian already has the $\mathfrak{su}(2)$ structure
\begin{equation}
    H_{\text{edge}}=\frac{J}{4} Z_{1}Z_{2}+\frac{\Omega(t)}{2} (\cos (\phi(t))X_{1}+\sin (\phi(t))Y_{1}) ).
    \label{eq:Hedge}
\end{equation}
Using $H_{\text{edge}}$ it is possible to do a $\frac{\pi}{2}$ ZZ rotation by simply waiting time $t=\pi/(4J)$. An identity is always possible by driving at strengths $\Omega=2\sqrt{(2 n \pi/t)^2-(\frac{J}{4})^2}$ for any integer $n$ which yields a real result. Finally, a $\frac{\pi}{2}$ $X$ rotation is possible by using an Euler angle decomposition using $\Omega_1=\Omega_3=-\Omega_2=J/2$ with
\begin{align}
     t_1=t_3&=\frac{\sqrt{2}\arctan(\sqrt{2})}{J}, \\ t_2&=\frac{5\sqrt{2}\pi }{3J}
 \end{align}
being the real solution with the shortest time.

\section{Smooth pulses using dynamical invariants}\label{sec:dynamicalinvariants}
The relatively long time for the $\frac{\pi}{2}$ $X$ rotation using square pulses is the reason the iSWAP takes $\approx 116/J$.  Instead of performing the $\pi/2$ $X$ rotations and $X$ gates using a sequence of square pulses, it is possible to perform smooth controls using the dynamical invariant of the Hamiltonian \cite{ChenMuga2011,LewisRiesenfeldPhase}. Using smooth pulses instead of square pulses is also preferable in practice, due to the limited bandwidth capabilities of arbitrary wave form generators \cite{Barnes2015,Machnes2018} and low-pass filters used to suppress the noise from the surrounding circuitry \cite{Chan2018a,VanDijk2019}.

In this section it will be shown that using this dynamical invariant approach the iSWAP gate can be implemented in $\approx19.5/J$ given a tolerance for a small error. A  dynamical invariant $I(t)$ obeys Eq.~\eqref{eq:DI}. 
For an $\mathfrak{su}(2)$ Hamiltonian with controls on two Pauli generators, such as Eq.~\eqref{eq:Hpp2} \eqref{eq:Hmm2}, the general form of $I(t)$ is known to be \cite{ChenMuga2011,Ralph2022dynamical}
\begin{equation}
    I(t) = \frac{\Omega_\text{ref}}{2} (\cos(\gamma) Z +\sin(\gamma) ( \sin(\beta) X+\cos(\beta)  Y)) \label{eq:dynamical0}
\end{equation}
where $\gamma$ and $\beta$ are the dynamical invariant parameters and $\Omega_\text{ref}$ is an arbitrary constant with units of energy. The  $\mathfrak{su}(2)$ generators $Z_{s_1 s_2}$ and $X_{s_1 s_2}$ are chosen to be equivalent to an $X$ and $Z$ Pauli matrices in Eq.~\eqref{eq:dynamical0}. This unconventional choice is made because it later simplifies the relation of $\gamma$ and $\beta$ to $\Omega$ and $J$, the Hamiltonian control parameters in $H_{s_1,s_2}$. Namely, this relation becomes
\begin{gather}
\dot{\gamma} = -J \sin(\beta) \label{eq:dynamical1}
\\
\Omega-\dot{\beta} = J \cot(\gamma)\cos (\beta)
\end{gather}
The evolution operator at the final time, $U(t_f)$, can be expressed in the initial and final values of $\gamma$, $\beta$ and the Lewis-Riesenfeld phase $\alpha$,
\begin{align}
   U(t_f)=& R_Z(-\beta(t_f))R_Y(-\gamma(t_f)) \times\\ \nonumber
   & R_Z(2\alpha(t_f)-\beta(t_f)+\beta(0))\times \\
   & R_Y(\gamma(0))R_Z(\beta(0)), \label{eq:Utfdynamical}\nonumber
\end{align}
where $R_A(\theta)$ is a rotation of angle $\theta$ around the $A$-axis.

By choosing $\beta=\arcsin(-\frac{\dot{\gamma}}{J})$ Eq.~\eqref{eq:dynamical1} is always solved since the exchange, $J$, is constant. Additionally, this means the other Hamiltonian control parameter, $\Omega$, is described by
\begin{equation}
    \Omega=\frac{-\ddot{\gamma}}{\sqrt{J^2-\dot{\gamma}^2}}+\cot(\gamma)\sqrt{J^2-\dot{\gamma}^2}.\label{eq:omegadynamical}
\end{equation}
The parametrization of $\Omega$ in Eq.~\eqref{eq:omegadynamical} is equivalent to that found in Ref.~\cite{Barnes2013,Barnes2015} where the dynamical invariants were not the focus.
Because the goal is to create $Z$, equivalent to $X_{s_1,s_2}$, rotations, we simplify the problem by choosing $\gamma$ to start and end at 0. By additionally setting $|\dot{\gamma}|$ to start and end at $J$, the evolution operator becomes $R_X(2\alpha(t_f))$. The Lewis-Riesenfeld phase can be expressed in terms of an integral shown from Ref.~\cite{Ralph2022dynamical},
\begin{equation}
    \alpha(t_f)=
\frac{1}{2}\int_0^{t_f} \frac{\cot(\beta)\gamma}{\sin(\gamma)}dt \label{eq:LewisRiesEq1}
\end{equation}
To create an $X$ gate in the $H_{++}$ and $H_{--}$ subspaces defined in Eqs.~\eqref{eq:Hpp2}-\eqref{eq:Hmm2} requires that $\alpha(t_f)$ be a multiple of $\pi/2$ while a $\pi/2$ $X$ rotation requires that $\alpha(t_f)$ is $\pi/4$, modulo $\pi$. 

\begin{figure}[!h]
    \centering
    \includegraphics[width=0.8\linewidth]{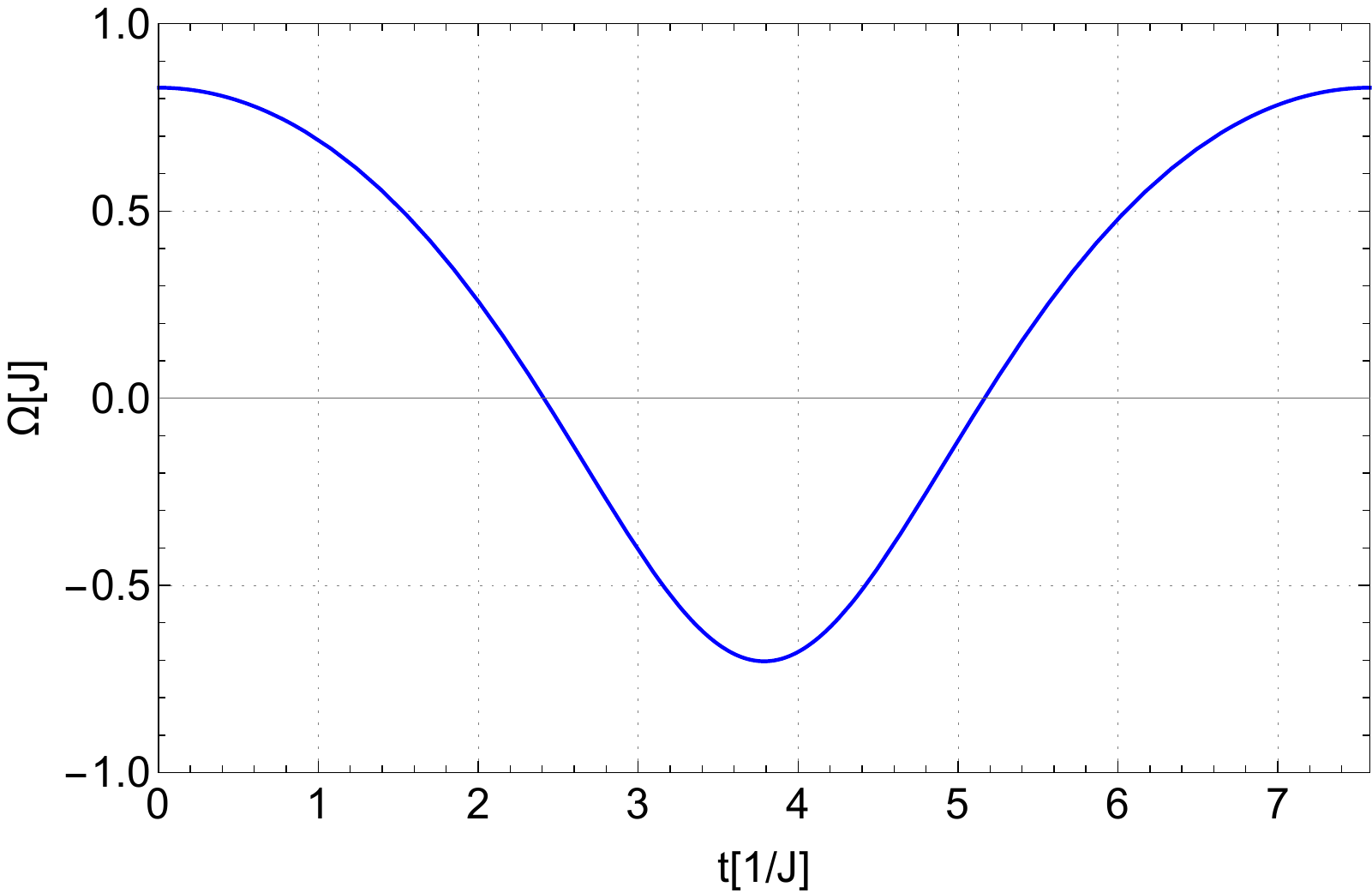}
    \caption{Pulse shape of the driving strength $\Omega$ needed to create an $X$ gate in the $H_{++}$ and $H_{--}$ subspaces.}
    \label{fig:2}
\end{figure}

\begin{figure}[!h]
    \centering
    \includegraphics[width=0.8\linewidth]{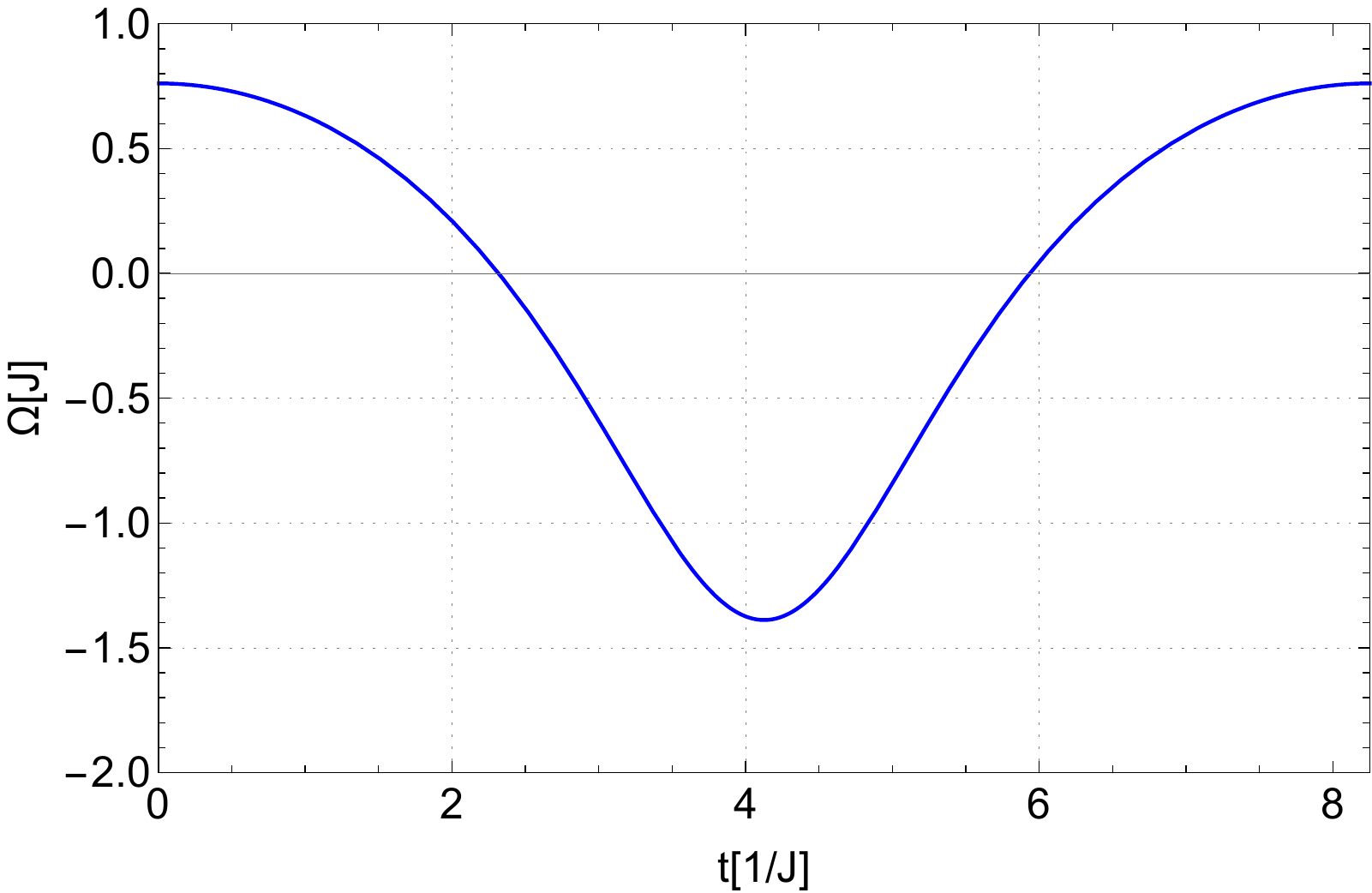}
    \caption{Pulse shape of the driving strength $\Omega$ needed to create a $\pi/2$ $X$ rotation in the $H_{++}$ and $H_{--}$ subspaces.}
    \label{fig:3}
\end{figure}

We satisfy the desired conditions $\gamma(0)=\gamma(t_f)=0$ and $|\dot{\gamma}(0)|=|\dot{\gamma}(t_f)|=J$ by choosing
\begin{equation}
    \gamma(t)=\frac{J}{\pi} \sin\left(\frac{\pi t}{t_f}\right)
    \label{eq:gammaoft}
\end{equation}
Additionally, the ansatz in Eq.~\eqref{eq:gammaoft} avoid the potential singularities in Eq.~\eqref{eq:omegadynamical} at intermediate times. 
Through numerical optimization an $X$ gate in the $H_{++}$ and $H_{--}$ subspaces was found at $t_f\approx7.58/J$ which results in the driving amplitude shown in Fig.~\ref{fig:2}. (The change in sign of the amplitude can be carried out in practice by a $\pi$ phase change.) The dynamics in the $H_{+-}$ and  $H_{+-}$ spaces can be corrected using an extra square pulse with the method from Sec.~\ref{subsec:Xgate}. The final square pulse has a strength $\Omega\approx 0.272 J$ and time $t\approx6.06/J$. This results in an infidelity of $10^{-8}$ which is close to the precision of the optimization used. A $\pi/2$ $X$ rotation is found by driving using Eq.~\ref{eq:gammaoft} with $t_f\approx8.25/J$ resulting in driving as shown in Fig.~\ref{fig:3} and using a final square pulse with $\Omega\approx1.41 J$ and $t\approx3.63/J$. If these gates are used instead of the square gates the total time for the iSWAP gate would be $t_\text{iSWAP}\approx 54.4/J$ instead of $\approx 116/J$ using purely square pulses.

\begin{figure}[!h]
    \centering
    \includegraphics[width=0.8\linewidth]{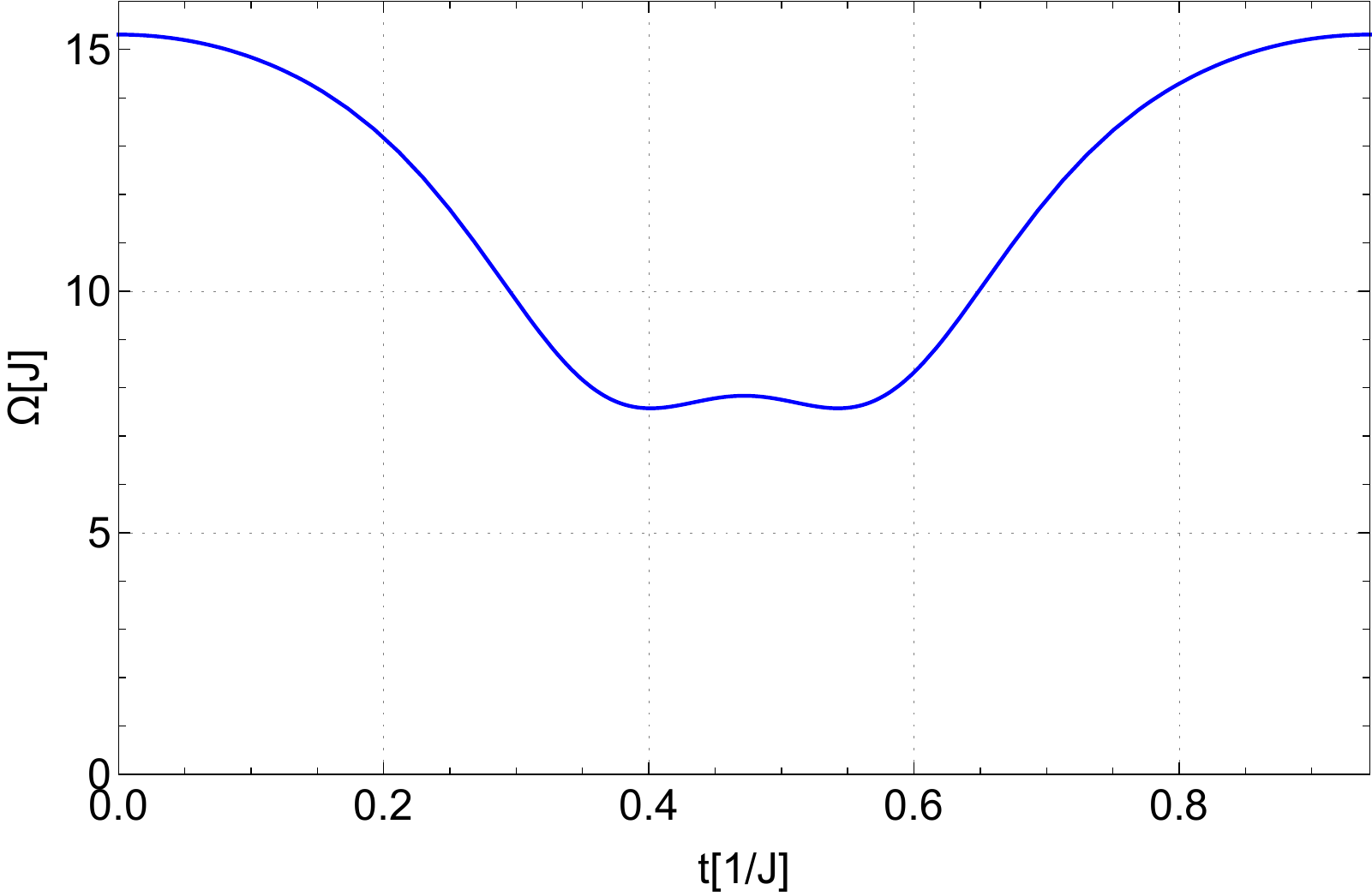}
    \caption{Pulse shape of the driving strength $\Omega$ needed to create a $\pi/2$ $X$ rotation up to a small error.}
    \label{fig:4}
\end{figure}

Performing a $\frac{\pi}{2}$ $X$ rotation using dynamical invariants without the extra square pulse to fix the $H_{+-}$ and $H_{-+}$ subspaces is possible when optimizing a function with more parameters. This can be achieved by choosing $\gamma(t)$ to have the ansatz 
\begin{equation}
    \gamma(t)=\left( \frac{J t_f}{\pi}-3 c_1\right) \sin\left(\frac{\pi t}{t_f}\right)+ c_1 \sin\left(\frac{3 \pi t}{t_f}\right). \label{eq:ansatzgamma2}
\end{equation}
The ansatz in Eq.~\eqref{eq:ansatzgamma2} guarantees that $\gamma$ starts and ends at 0 and $|\dot{\gamma}|$ starts and ends at $J$. Additionally, as long as $|\dot{\gamma}|\leq J$ and $\gamma$ does not reach $n \pi$ when $|\dot{\gamma}| \neq J$ for any integer $n$ a valid solution will be found \cite{Barnes2015}. A single-shot $\frac{\pi}{2}$ $X$ rotation requires not only $\alpha(t_f)=\pi/2+n\pi$ as well as $\int_0^{t_f} \Omega(t)dt=\pi/2+m\pi$ for any integers $m$ and $n$ as longs as both are odd or both are even. At $c_1\approx0.0537$ and $t_f\approx0.944/J$ a solution is found that is close to a $\frac{\pi}{2}$ $X$ rotation.  The resulting pulse shape for $\Omega$ is shown in Fig.~\ref{fig:4}. It is not an exact solution as it has a state-averaged infidelity $1-F\approx1.5 \times 10^{-4}$. The average fidelity is calculated using $F=\frac{1}{9}+\frac{1}{576}\sum_{i,j,k=I,X,Y,Z} \text{Tr}(U_{t}\sigma_{ijk}U^{\dagger}_{t}\mathcal{M}(\sigma_{ijk}))$ where $U_t$ is the target evolution, $\sigma_{ijk}=\sigma_i \otimes\sigma_j \otimes\sigma_k$ is the Kronecker product of Pauli matrices and $\mathcal{M}$ is the trace preserving map that time evolves the density matrix $\rho(t) = \mathcal{M}(\rho(0))$ \cite{Cabrera2007}. Although it is not an exact solution, this method is 20 times faster than using square pulses, taking only $t_{f}\approx0.944/J$. The speed-up is partly because this solution enables the use of a higher maximum driving of $\Omega_\text{max}\approx 15.3 J$ and partly because it does not require an extra square pulse to fix the evolution caused by $H_{+-}$ and $H_{-+}$. An $X$ gate can be created by doing this pulse shape twice. Using this method to create an iSWAP gate would reduce the time down to $t_\text{iSWAP}\approx 13.8/J$ time and would have a trace infidelity of  $1-F\approx 2.4 \times 10^{-3}$.

Multi-qubit gate fidelities are mainly limited by charge noise though, which causes fluctuations in the voltage \cite{Throckmorton2017a,Buterakos2021,Culcer2009,Yoneda2017,Xue2022,Yang2019b,Connors2022,Takeda2013}. As a result, the exchange strength, $J$, fluctuates in the presence of charge noise because it depends on the voltage. To analyze the effect of exchange fluctuations on this iSWAP implementation, each type of section in the sequence will be examined individually. Numerically calculating the average infidelity of the $\frac{\pi}{2}$ $X$ rotation due to quasistatic exchange fluctuations, $\delta J$, we find that the infidelity is approximately $(1.5 +16 (\delta J/J)^2) \times 10^{-4}$ for $\delta J/J < 2$. The infidelity of a simultaneous identity performed using a square pulse is negligible, as previously noted in Ref.~\cite{Kanaar2021}, approximately $(3.5 +3.1(\delta J/J)^2+263(\delta J/J)^4 )\times 10^{-7}$ within the same range of exchange fluctuations. The most error-prone component of the sequence is the $\frac{\pi}{2}$ $ZZ$ rotations, which require waiting for $T=\pi/J$, and have an average infidelity of approximately $1.1(\delta J/J)^2$ for $\delta J/J < 0.25$.
If an iSWAP gate in a three qubit device is performed as described at the end of Section \ref{subsubsec:ZZrot} then a total of eight $\frac{\pi}{2}$ $X$ rotations, two $\pi/J$ waiting periods and three identities have to be performed.   
In the worst case, all these infidelities of all these operations add together, giving an upper bound on average infidelity of $1.4\times 10^{-3}$ at 1\% quasistatic fluctuation in exchange, $\delta J/ J = 0.01$. 
Another source of error could be unwanted next-nearest neighbor coupling. However, next-nearest neighbor coupling in a chain of quantum dots filled with one electron is predicted to be vanishingly small, and is usually neglected \cite{Chan2021}.


\section{Expanding beyond a linear chain}
\begin{figure}
    \centering
    \includegraphics[width=0.8\linewidth]{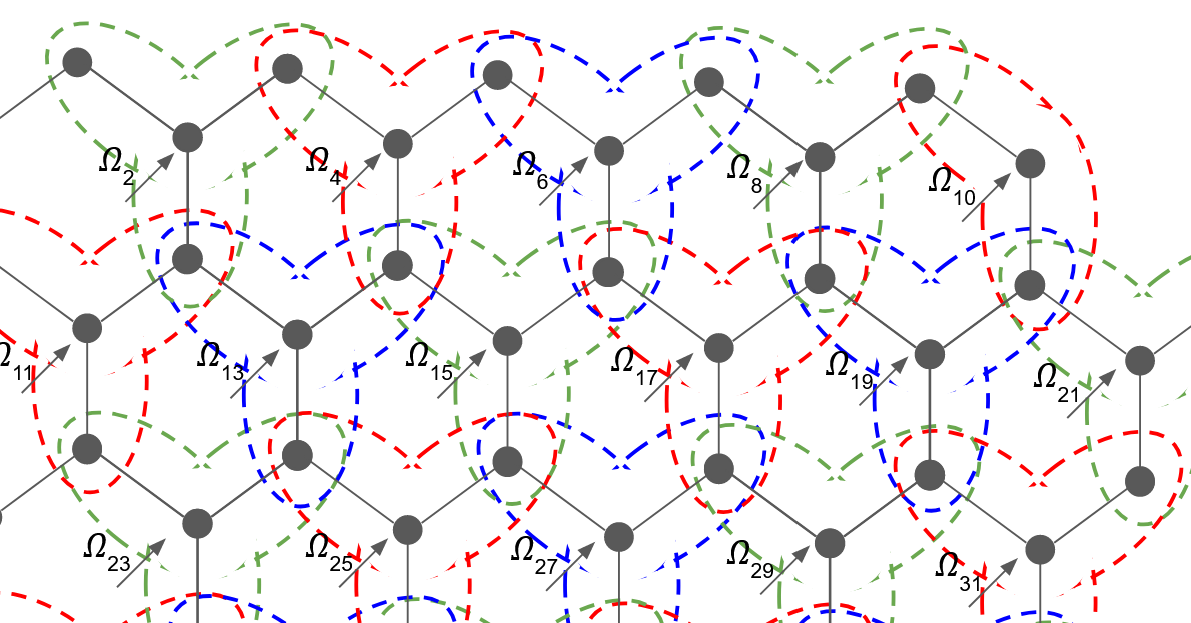}
    \caption{Schematic representation of the driving on 2D honeycomb array of qubits which results in a decomposition of $\mathfrak{su}(2)$s. The dashed ``Y" shapes indicate the spin subsets involved in the separate $\mathfrak{su}(2)$s.}
    \label{fig:6grid2d}
\end{figure}

The decomposition of the linear chain Hamiltonian in Eq.~\eqref{eq:HRN} into a set of $\mathfrak{su}(2)$s and $\mathfrak{u}(1)$s works for a larger number of nearest neighbors as well. The simplest example beyond a linear chain is a 2D honeycomb lattice, shown in Fig.~\ref{fig:6grid2d}, which has $M=3$ nearest neighbors. When driving every other lattice point as indicated in Fig.~\ref{fig:6grid2d} the Hamiltonian of the terms around the driven lattice points, as indicated by the colored dashed lines, forms sets of commuting $\mathfrak{su}(2)$s again. 
In this case, the Hamiltonian of one dashed region centered on spin $c$ decomposes into 8 $\mathfrak{su}(2)$s,
\begin{equation}\label{eq:8su2s}
    H_c = \sum_{s_1 = \pm}\sum_{s_2 = \pm}\sum_{s_3 = \pm}H_{s_1 s_2 s_3}
\end{equation}
where
\begin{equation}\label{eq:honeycombH}
    H_{s_1 s_2 s_3} = \left(\sum_{i=1}^3 s_i \frac{J_{c,nn(c,i)}}{4}\right) Z_{s_1 s_2 s_3} + \left(\prod_{i=1}^3 s_i\right) \frac{\Omega_c}{2} X_{s_1 s_2 s_3},
\end{equation}
with $nn(c,i)$ the label of the $i$th of the three nearest neighbors of central spin $c$, and, in analogy to Eqs.~\eqref{eq:eff_Z}-\eqref{eq:eff_X}, 
\begin{equation}\label{eq:honeycombZ}
    Z_{s_1 s_2 s_3} = Z_c\prod_{i=1}^3\frac{1}{2}\left(Z_{nn(c,i)} + s_i I\right)
\end{equation}
and equivalently for $X_{s_1 s_2 s_3}$. Fortunately, when setting $J_{c,nn(c,i)}=J$ for all $i$, the 8 $\mathfrak{su}(2)$ Hamiltonians of Eq.~\eqref{eq:honeycombH} still contain just two unique sets of coefficients up to the signs, corresponding to either all $s_i$ being the same or one of them being different, as in Eqs.~\eqref{eq:Hpp2}-\eqref{eq:Hmm2}. So, there are only two sets of constraints to satisfy for each of the local gates, the same as in Sec.~\ref{sec:Individualgates}. This reduction is only possible for $M<4$.

Unlike in Sec.~\ref{sec:Individualgates}, none of the $\mathfrak{su}(2)$s simplify to $\mathfrak{u}(1)$s, so the pulse design is more tedious. However, using this decomposition and numerically searching over the parameters of a symmetric five-piece sequence of square pulses one can find rotations of the central spin about $X$ by $\pi/2$ or $\pi$ as given in Table \ref{tab:honeycomb} by $X_{\pi/2}$ and $X_{\pi}$. Simultaneous identities on other lattice points can be achieved by the pulse sequences in Table \ref{tab:honeycomb} indicated by $ \sqrt{I_{\pi/2}}$ and $ \sqrt{I_{\pi}}$ for $\pi/2$ and $\pi$ $X$ rotations. 
\begin{table}\label{tab:honeycomb}
\centering
\begin{tabular}{ |c||c|c|c|c|c|c||  }
 \hline
  & $\Omega_1=\Omega_5$ & $t_1 = t_5$ & $\Omega_2 = \Omega_4$ & $t_2 = t_4$ & $\Omega_3$ & $t_3$ \\
 \hline
 \hline
 $X_{\pi/2}$ & 0.5 J & 9.77/J & 0 & 2.57/J & 1.40 J & 1.93/J \\
 $X_{\pi}$ & 0.5 J & 9.51/J & 0 & 4.38/J & -2.16 J & 1.22/J\\
$ \sqrt{I_{\pi/2}}$ & 1.74 J & 2.38/J & 0 & 3.23/J & 1.13 J & 2.10/J\\
 $\sqrt{I_{\pi}}$ & 1.22 J & 3.16/J & 0 & 2.33/J & 0.67 J & 3.52/J\\
 \hline
\end{tabular}
\caption{5-piece pulse sequence parameters for local rotations of a spin in a honeycomb lattice with fixed couplings. The $\sqrt{I}$ sequences are to be performed twice to produce an identity with the same duration as the corresponding $X$ rotation.}
\end{table}
The iSWAP gate can be formed by interspersing these rotations on the central spins of the subsets in between undriven entangling periods just as in Sec.~\ref{sec:iswapPLUSindividualgates}. The total time for the whole iSWAP sequence using these pulses is $\approx 195/J$. 

It would be desirable to greatly reduce this time using faster smooth pulses instead of the proof-of-principle square pulse sequence above, as in Sec.~\ref{sec:dynamicalinvariants}. There seems to be no reason this should not work, but note that it is a substantially more cumbersome calculation now since, as noted above, there are two types of $\mathfrak{su}(2)$ Hamiltonian arising from Eq.~\eqref{eq:honeycombH}. Their constant $Z$ coefficients differ by a factor of 3 while their controllable $X$ coefficients must be identical. Thus, using the dynamical invariant approach requires one to find two functions, $\gamma_1$ and $\gamma_2$, such that, from Eq.~\eqref{eq:omegadynamical},
\begin{align} \label{eq:pairedgammas}
    &\frac{-\ddot{\gamma_1}}{\sqrt{J^2-\dot{\gamma_1}^2}}+\cot(\gamma_1)\sqrt{J^2-\dot{\gamma_1}^2} \\ 
    =& \frac{-\ddot{\gamma_2}}{\sqrt{9J^2-\dot{\gamma_2}^2}}+\cot(\gamma_2)\sqrt{9J^2-\dot{\gamma_2}^2}.\nonumber
\end{align}
while also obeying the boundary conditions $\gamma_i(0) = \gamma_i(t_f) = 0$, $|\dot{\gamma}_1(0)| = |\dot{\gamma}_1(t_f)| = J$, $|\dot{\gamma}_2(0)| = |\dot{\gamma}_2(t_f)| = 3J$ and producing the desired Lewis-Riesenfeld phases. We have not yet found a good way to enforce Eq.~\eqref{eq:pairedgammas} for general smooth functions, although the square pulse solution found above is a specific example. This is a topic for future investigation.

Furthermore, we note that the decomposition approach of this section works for any lattice configuration as long as the set of nearest neighbors is disjoint with the set of next-nearest neighbors. The number of $\mathfrak{su}(2)$s will depend on the configuration and the relative values of $J_{ij}$. A lattice point with $M$ connections when driven resonantly will create a possible $2^M$ $\mathfrak{su}(2)$s which commute with all $\mathfrak{su}(2)$s created by driving on any another lattice point that is not a nearest neighbor. The form of the Hamiltonians can be found by extending Eqs.~\eqref{eq:8su2s}-\eqref{eq:honeycombZ} to a larger number of nearest neighbors. Some of the $\mathfrak{su}(2)$s might simplify to $\mathfrak{u}(1)$s as in the case of the linear chain in Sec.~\ref{sec:Model4} when the coupling $J$ is identical for all nearest neighbor pairs. Generally, though, the problem becomes one of designing a pulse that carries out the desired evolution in each of the $\mathfrak{su}(2)$s at the same time, as in Sec.~\ref{subsec:Xgate}. A large number of $\mathfrak{su}(2)$s then corresponds to having many constraints that the pulse must satisfy.
This can be seen from examining a square lattice ($M=4$) with identical couplings. The Hamiltonian decomposes into 16 $\mathfrak{su}(2)$s, of which \emph{three} contain a unique set of coefficients modulo the signs. The simple square pulse form of Sec.~\ref{sec:Individualgates} cannot work because there are more constraints than free parameters. Using a longer sequence of square pulses (as we did above for the honeycomb lattice) or reverse-engineering a smooth pulse using dynamical invariants as in Sec.~\ref{sec:dynamicalinvariants} is necessary, and the numerics become more challenging. Importantly, though, the difficulty of the problem does not depend on the size of the lattice, only its connectivity. The algebraic form of the decomposition is independent of the size of the lattice, and so is the time required to perform an iSWAP on a desired link.

\section{Conclusion}
In summary, we have used Cartan decomposition and dynamical invariants to create an iSWAP gate in a chain of always-on exchange coupled qubits which can be used for spin state transfer. The Cartan decomposition in the chain of always-on exchange coupled qubits is found by driving every other qubit. This results in sets of four $\mathfrak{su}(2)$ Hamiltonians for every driven qubit. 
To implement an iSWAP gate in these Hamiltonians only $\frac{\pi}{2}$ $X$ rotations, $\frac{\pi}{2}$ $ZZ$ rotations, and $X$ gates are required if virtual $Z$ gates are possible. Using square pulses and Euler angle decomposition, it was shown that these gates can be implemented. This method of creating an iSWAP gate takes $t_\text{iSWAP}\approx 116/J$. To do this faster, a method using the dynamical invariants to create smooth fields to create an iSWAP gate was shown (we note that the obtained dynamical invariants can in principle be used for other purposes, such as transitionless quantum driving of a single state). This yielded a iSWAP gate in $t_\text{iSWAP}\approx 54.4/J$ for an analytical solution or in only $t_\text{iSWAP}\approx 13.8/J$ for a numerical solution with an infidelity of $2.7\times10^{-3}$. 
The decomposition into $\mathfrak{su}(2)$s also works for a 2D array of qubits with more than $2$ nearest neighbors as long as the set of nearest neighbors is disjoint from the set of next-nearest neighbors. In the case of a honeycomb lattice, we have shown how to perform directed transport of the spin state on the lattice without control of the coupling. For other lattices, solutions are expected to exist, but exploring the methods to find them is an open task for future research.\vskip6pt

\section*{Acknowledgements} \label{sec:acknowledgements}
This material is based upon work supported by the Army Research Office (ARO) under Grant Number W911NF-17-1-0287, and by the National Science Foundation under Grant No. 1915064.

\bibliographystyle{apsrev4-1}
\bibliography{refs}
\end{document}